\documentclass[prl, twocolumn,showpacs]{revtex4}
\usepackage{amssymb}
\usepackage{amsmath}
\usepackage{graphicx}
\makeatletter

\begin{document}

\title{Theory for Superconductivity in Iron Pnictides at Large Coulomb U Limit}
\author{Wei-Qiang Chen$^{1}$, Kai-Yu Yang$^{2,1}$, Yi Zhou$^{3}$, and Fu-Chun Zhang$^{1,4}$
}
\affiliation{$^{1}$Department of Physics, and Center of Theoretical and Computational
Physics, the University of Hong Kong, Hong Kong, China\\
$^2$Institut fur Physik, ETH - Zurich, 8093 Switzerland\\
$^3$Department of Physics, Chinese University of Hong Kong, Hong Kong, China\\
$^4$Department of Physics, Zhejiang University, Hangzhou, China}
\date{\today}

\begin{abstract}
Superconductivity in iron pnictides is studied by using a
two-orbital Hubbard model in the large U limit.
The Coulomb repulsion induces an orbital-dependent pairing
between charge carriers.  The pairing is found mainly from the 
scattering within the same Fermi pocket.  The inter-pocket pair scatterings
determine the symmetry of the superconductivity, which is
extended s-wave at small Hund's coupling, 
and d-wave at large Hund's coupling
and large U. The former is consistent with recent experiments of
ARPES and Andreev reflection spectroscope.
\end{abstract}


\maketitle

Superconducting (SC) iron pnictides have the
highest transition temperature next to the
cuprates\cite{Kamihara,XHChen,NLWang,HHWen,Zhao,122,Xu}. The
parent compounds are metallic  spin density wave (SDW) state
~\cite{PDai,Mandrus,LDA,Fang}. 
Superconductivity occurs when part of $\mathrm{Fe}^{2+}$ ions are replaced
by $\mathrm{Fe}^{+}$. A multi-orbital Hubbard model may be a starting point
to study the superconductivity.~\cite{Kotliar,Mazin,Kuroki,Dai,FWang,
Yao,SCZhang,Si,DHLee,PALee,Baskaran}
Since the parent compound is metallic,
most theories examine the SC instability
from weak Coulomb interaction point of view
\cite{Mazin,Kuroki,FWang,Yao,DHLee,PALee}. 
On the other hand, the observed magnetic moment in the SDW phase is large
\cite{neutron}, indicating importance of spin couplings.
The dynamic mean 
field theory \cite{Kotliar} also suggests its closeness to a Mott insulator. 
This calls for an alternative approach from the viewpoint of
large Coulomb repulsion U, which 
will be the purpose of the 
present letter.

The electronic states of the compound 
are predominantly Fe-3d orbitals near the 
Fermi surface (FS) \cite{LDA, Fang, Boeri}, which is
comprised of two hole pockets centered at $\Gamma=(0,0)$ and two 
electron pockets at $X=(\pi,0)$ and $Y=(0,\pi)$, 
in the unfolded Brillouin zone (BZ), corresponding to
1 Fe atom per unit cell. 
Note that the buckling of As-atoms reduces the BZ to the square enclosed by the 
dashed lines in Fig. 1. 
The FS structure can be reproduced by a 
5-orbital model \cite{Kuroki}. 
The bands near the FS are mainly 
$d_{xz}$ and $d_{yz}$ orbitals \cite{Boeri}, and the FS in the reduced BZ can be reproduced by a 2-orbital model, which
shifts a hole Fermi pocket from the $\Gamma$- 
to the $M=(\pi,\pi)$-points in the unfolded BZ. In this letter, we use the 2-orbital model to study the superconductivity 
at large U limit.  We argue that our qualitative results will remain unchanged due to 
the simplification of the 2-orbital model. We find that 
the virtual hopping induces orbital dependent pairings of charge carriers.
The intra Fermi pocket pair scattering is strongest, and 
the pairing symmetry is determined by inter pocket pair scatterings and is extended s-wave 
($s_{\pm})$ for small Hund's coupling
and d-wave for large Hund's coupling and large U. The $s_{\pm}$- state
was proposed by Mazin et al. ~\cite{Mazin} based on the analysis of 
the small Fermi pockets and spin fluctuations, and was found in weak coupling or small U 
approaches \cite{Kuroki,FWang}. Our result appears 
consistent with the ARPES~\cite{Ding} and Andreev 
reflection spectroscope~\cite{Chien}. 

The 2-orbital model reads\cite{SCZhang}
$H= H_0 + H_I$, where  $H_I$ is an on-site Coulomb term, and $H_0$ is a tight-binding
model on a square lattice of Fe- atoms,
\begin{eqnarray}
  H_{0}=\sum_{\mathbf{k}nm\sigma}(\epsilon^{nm}_{\mathbf{k}}-\mu)
\hat{c}^{\dag}_{\mathbf{k}n\sigma}\hat{c}_{\mathbf{k}m\sigma}
=\sum_{\mathbf{k}\alpha\sigma}\xi_{\mathbf{k}\alpha} \hat{c}^{\dag}_{\mathbf{k}\alpha\sigma}
\hat{c}_{\mathbf{k}\alpha\sigma},
\label{EQ:HTB}
\end{eqnarray}
where $\epsilon^{nm}_{\mathbf{k}}$ is the hopping matrix in $\mathbf{k}-$ space,
$n=1$ or $2$ denote orbitals $d_{xz}$ (or $d_{yz}$). $\mu$ is the chemical potential.
$\alpha=\pm $ represents the electron  or upper ($+$) band and the  hole  or lower ($-$) band,
 corresponding to the diagonalized energy $\xi_{\mathbf{k} \pm}$. The
band and orbital representations are related by a
unitary transformation, $\hat{c}_{\mathbf{k}n\sigma}=\sum_{\alpha
= \pm} u_{n\alpha}(\mathbf{k})\hat{c}_{\mathbf{k} \alpha \sigma}$.
Here we follow Ref. \cite{SCZhang} and parameterize $H_0$ by 
hopping integrals $t^{nm}_{\vec{\tau}}$ between two sites $i$ and
$j=i+\vec{\tau}$, which is the Fourier transform of
$\epsilon_{nm}(\mathbf{k})$. We set
$t^{11}_{\hat{x}}=t^{22}_{\hat{y}}=t_1$,
$t^{11}_{\hat{y}}=t^{22}_{\hat{x}}=t_2$, $t^{nn}_{\hat{x} \pm
\hat{y}} =t_3$, and $t^{12}_{\hat x \pm \hat y} = \pm t_4$ by lattice and orbital symmetry.

\begin{figure}[htbp]
\centerline{\includegraphics[width=0.3\textwidth]{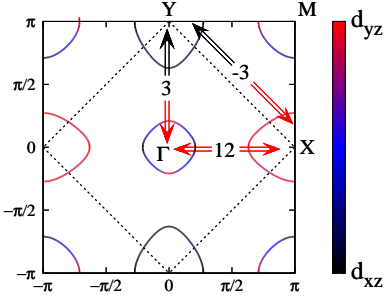}}
\caption[]{\label{fig:fs} (Color online) Fermi surface in the unfolded Brillouin zone (BZ)
of  $H_0$. The square enclosed by the dashed lines is the reduced BZ.
 Color scheme
  illustrates weights contributed from orbitals $d_{xz}$ and $d_{yz}$. Arrows indicate 
inter-pocket pair scatterings with wavevectors $\mathbf{q}\sim(0,\pi)$, 
$(\pi,0)$ and $(\pi,\pi)$.  
Numerics (positive value: attractive)
 are the corresponding scattering 
amplitudes $A_{nn}^{mm}( \mathbf{q})$ 
in Eq. (7) in unit of $t^2/U$ at $J=0$. Not shown is the intra-pocket scattering 
$A_{nn}^{nn}(\mathbf{q} \sim (0,0)) = 20 t^2/U$. 
}
\end{figure}

By choosing $t_1 = -t$, $t_2=1.3t$, $t_3=t_4=-0.85t$,
the calculated FS with electron density per site $\approx 2.10$ is
reproduced in Fig.~\ref{fig:fs}, which is similar to the first principle
calculations\cite{Mazin,SCZhang} for LaFeAsO. 
The weight contributed from each orbital at the FS is
illustrated in the figure. The state on the
electron pocket around the $X$ ($Y$) is mainly from $d_{yz}$ ($d_{xz}$) orbital.
The state on the hole pocket around the $\Gamma$ consists of $d_{yz}$ and $d_{xz}$ orbitals
equally if $\bf k$ is along the diagonals, and mainly from $d_{xz}$ (or $d_{yz}$) orbital if along 
the $x$ or $y$ axis.

The on-site interaction
\begin{eqnarray}
H_{I}=\sum_{i;m=1,2}[U \hat{n}_{im\uparrow}\hat{n}_{im\downarrow}+
J\hat{c}^{\dag}_{im\uparrow}\hat{c}^{\dag}_{im\downarrow}\hat{c}_{i\overline{m}\downarrow}
\hat{c}_{i\overline{m},\uparrow}] \notag \\
+\sum_{i;\sigma \sigma\prime} [U_{12}\hat{n}_{i1\sigma}\hat{n}_{i2\sigma\prime}
+J\hat{c}^{\dag}_{i1\sigma}\hat{c}^{\dag}_{i2\sigma\prime}
\hat{c}_{i1\sigma\prime}\hat{c}_{i2\sigma}]
\label{EQ:HI}
\end{eqnarray}
where $\hat n_{im\sigma}=\hat c^{\dag}_{im\sigma}\hat
c_{im\sigma}$, $U$ and $U_{12}$ are the intra- and
inter-orbital direct Coulomb repulsions, respectively. The terms with $J$
are the exchange interaction.
By symmetry, $U = U_{12}+ 2J$. \cite{Castellani}
In the limit, $U >> t$, 
each lattice site is doubly occupied in the parent compound.
Upon electron doping, some sites will have 3 electrons (or 1 hole).
A single hole at site $i$ may interchange with a two-hole state at site $j$,
leading to a metallic phase.
The effective interaction between two
single holes on neighboring sites ($i$,$j$) can be derived by using
second order perturbation theory, and it is given by
\begin{eqnarray}
  H_2&=&-\sum_{ij}\sum_{nmn'm'} \Bigl[ A_{nm}^{m'n'}(ij)\hat{b}^{\dag}_{nm}(ij)\hat{b}^{n'm'}(ij)
  \notag \\
  &&\phantom{-\sum_{ij}} +\sum_{S_{z}}B_{nm}^{m'n'}(ij)\hat{T}^{S_{z}\dag}_{nm}(ij)\hat{T}^{n'm'}_{S_{z}}(ij)\Bigr]
\label{EQ:Heff}
\end{eqnarray}
where $S_z=-1, 0, 1$, and
\begin{eqnarray}
A_{nm}^{m'n'}(ij)&=&[\frac{(-1)^{m+m'}}{U-J}+\frac{1}{U+J}]t_{ij}^{nm}t_{ji}^{m'n'}
+\frac{t_{ij}^{n\bar{m}}t_{ji}^{\bar{m'}n'}}{U_{12}+J} \notag \\
B_{nm}^{m'n'}(ij)&=&\frac{(-1)^{m+m'}}{U_{12}-J}t_{ij}^{n\bar{m}}t_{ji}^{\bar{m'}n'},
\label{EQ:ABcoef}
\end{eqnarray}
where $\bar{m}$ refers to the conjugate orbital of $m$, and
the 1st and 2nd terms in $H_2$ are the pairing
interactions in the spin singlet and triplet channels,
respectively. The spin singlet pair operator
$\hat{b}^{nm}(ij)=\frac{1}{\sqrt{2}}(\hat{c}_{in\uparrow}
\hat{c}_{jm\downarrow} -
\hat{c}_{in\downarrow}\hat{c}_{jm\uparrow})$, and the spin triplet
pair operators $T_{S_z}$ can be written similarly. 
In Eq. (3) and formalism hereafter, we use hole notation.
The results plotted in all the figures, however,
will be in the electron convention.  Castellani \textit{et al.} \cite{Castellani} studied the spin-spin coupling for a
2-fold orbital degenerate Hubbard model in the context of $V_2O_3$.
Our expression here is equivalent to theirs, although the pairing forms were not explicitly given
in their formalism.
The spin triplet states become
important at $J/U \rightarrow 1/3$, or $J \rightarrow U_{12}$,
which can be seen clearly from
the term in $B$. Below we focus on 
the spin-singlet state with even parity, which is energetically
more favorable for $J/U$ not so large.  The pairing
interaction between carriers derived in the large U-
limit should be relevant to the intermediate coupling region
~\cite{Gossamer}.

The effective Hamiltonian is then $H_{eff}=H_0 + H_2$, subject to
the constraint of no more than 2 holes per site. This can
formally be represented by a Gutzwiller projection operator to
project all the unphysical states, similar to that in the t-J
model~\cite{Anderson}. 
$H_{eff}$ may be studied by using a renormalized
Hamiltonian approach to take into account the
projection~\cite{Zhang88} by introducing renormalization factors,
$g_t$ for $H_0$ and $g_2$ for $H_2$, both are doping
dependent. For a given doping, the effect of the renormalization is 
to scale all the $t's$ to $g_t t's$, and $(U,J)$ to $(g_t^2/g_2)(U,J)$.
Below we will absorb these
renormalization factors into the parameters ($t's$ and $U$) and
effectively set $g_t=g_2=1$ in our calculations.

\begin{figure}[htbp]
\centerline{\includegraphics[width=0.4\textwidth]{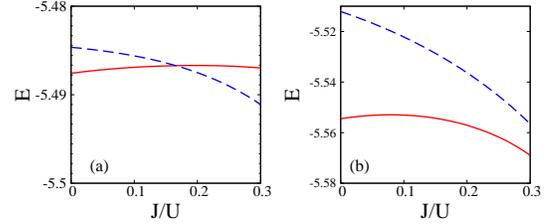}}
\caption{\label{fig:energy} (Color online) Energy per site of $H_{eff}$
in $s_{\pm}$ state  (red solid line) and d-wave state
(blue dashed line) for (a): $t/U=0.1$ and (b): $t/U=0.2$.
}
\end{figure}

$H_{eff}$ can then be solved using a mean field theory by 
introducing mean fields for the
spin-singlet pairing with even parity and symmetric orbitals \cite{Yi-Zhou},
$\Delta_{nm}(\vec{\tau})=\Delta_{mn}(\vec{\tau}) 
=\frac{1}{\sqrt{2}}\left\langle \hat{b}^{nm}(i,i+\vec{\tau})
\right\rangle$, with $\vec {\tau}=\pm \hat x, \, \pm \hat y, \, \pm (\hat x \pm \hat y)$.
By symmetry, depending on $s_{\pm}$ ($A_{1g}$) or d-wave
($B_{1g}$) states,
we have
$\Delta_{11}(\hat x) = \pm \Delta_{22}(\hat y)$,
$\Delta_{11}(\hat y) = \pm \Delta_{22}(\hat x)$,
$\Delta_{12}(\hat x) = \Delta_{12}(\hat y) = 0$,
$\Delta_{11} (\hat x \pm \hat y) = \pm \Delta_{22}(\hat y \mp \hat x)$,
$\Delta_{12}(\hat x + \hat y) = - \Delta_{12}(\hat x - \hat y)$. Note that
$\Delta_{12}(\hat x\pm \hat y)=0 $ for the d-wave state.
The pairing strength with $A_{2g}$ and $B_{2g}$ symmetries
~\cite{Yi-Zhou} are found very tiny, and will not be discussed further.\cite{dagotto}
The mean field Hamiltonian of $H_{eff}$ can be written as
\begin{align}
\label{EQ:Hband}
H_{MF} = \sum_{\mathbf{k}} \hat{\psi}_{\mathbf{k}}^{\dag}
\left( \begin{matrix}
  \xi_{\mathbf{k}} & V(\mathbf{k})\\
  V^{\dag}(\mathbf{k}) & - \xi_{\mathbf{k}}
\end{matrix} \right) \hat{\psi}_{\mathbf{k}},
\end{align}
where $\hat{\psi}^{\dag}_{\mathbf{k}} = \left( \hat{c}^{\dag}_{\mathbf{k}+\uparrow},
\hat{c}^{\dag}_{\mathbf{k}-\uparrow}, \hat{c}_{-\mathbf{k} +
\downarrow}, \hat{c}_{-\mathbf{k} - \downarrow}
\right)$. $V(\mathbf{k})$ is a $2 \times 2 $ matrix in band picture, given by
\begin{eqnarray}
V_{\alpha \beta}(\mathbf{k}) = \sum_{nmm'n';\tau} A_{nm}^{m'n'}(\vec{\tau})
\Delta^{*}_{nm}(\vec{\tau})e^{i \mathbf{k} \cdot \vec{\tau}}
u_{m'\alpha}(\mathbf{k})u_{n'\beta}(\mathbf{k}) \notag
\end{eqnarray}
$H_{MF}$ can be solved self-consistently, and 
the energy per site is  $E
=-\frac{1}{N}\sum_{\mathbf{k},\pm} E_{\pm}(\mathbf{k})$,
with $E_{\pm}(\mathbf{k})$ the
quasi-particle energy of the upper ($+$) and lower
($-$) bands, given by
\begin{eqnarray}
E_{\pm}(\mathbf{k})= \sqrt{ w^2_{+} +V_{+-}^2 \pm
\sqrt{w^4_{-} +V_{+-}^2[(\delta\xi)^2 + 4 \bar
{V}^2]}}
\end{eqnarray}
where $\delta\xi=\xi_{+}-\xi_{-}$, $\bar{V}=[V_{++}+V_{--}]/2$, and
$w^2_{\pm} =[\xi_{+}^2 + V^2_{++} \pm (\xi_{-}^2 + V^2_{--})]/2$,
and the $\mathbf{k}$-dependence is implied.
 In Fig.~\ref{fig:energy}, the energies of the SC states 
 are depicted as functions of $J/U$ for $t/U =
0.1$ and $t/U = 0.2$. At $t/U=0.2$, the $s_{\pm}$
 state is always energetically favorable.  
At $t/U= 0.1$, the ground state is $s_{\pm}$-wave if 
 $J/U < 0.16$ and a d-wave if $J/U > 0.16$.

\begin{figure}[htbp]
\centerline{\includegraphics[width=0.4\textwidth]{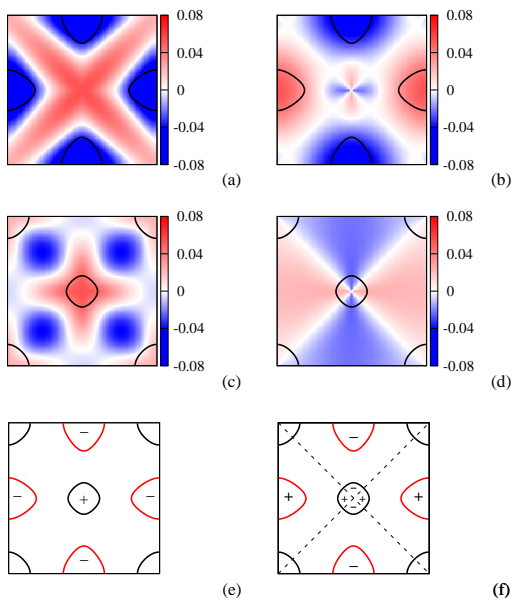}}
\caption[]{\label{fig:pairing} (Color online) Intra-band pairing amplitude
$V_{\alpha,\alpha}$ for $s_{\pm}$ (left) and
d-wave (right) symmetry states. Upper panels (a,b):
electron band, middle panels (c,d): hole band. Fermi surfaces are
indicated by the black lines. Lower panels (e,f): relative sign of
the pairing amplitudes and order parameters around Fermi pockets.
}
\end{figure}

In Fig. ~\ref{fig:pairing}, we plot the intra-band pairing amplitude 
$V_{++}(\mathbf{k})$ for the electron-band and $V_{--}(\mathbf{k})$
for the hole band.  In the $s_{\pm}$ state, $V(\mathbf{k})$ is invariant
under a $\pi/2$ rotation, and
$V_{++}(\mathbf{k})$
and $V_{--}(\mathbf{k})$ have a nodal line in the BZ. $V_{++}$
have the same sign on $X$ and $Y$ pockets, but are opposite to $V_{--}$ on $\Gamma$.
In the d-wave state,
$V_{\alpha\alpha}(\mathbf{k})$
changes a sign under a $\pi/2$ rotation, and has nodal lines
along the diagonals in the BZ. 

\begin{figure}[htbp]
\centerline{\includegraphics[width=0.35\textwidth]{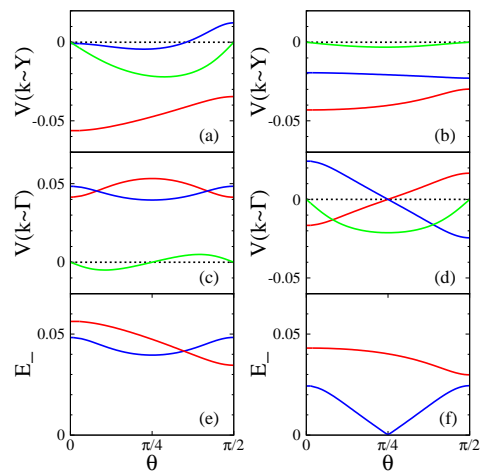}}
\caption[]{\label{fig:gap} (Color online) Angle dependence
of pairing amplitude $V_{++}(\mathbf{k})$ (red line),
$V_{--}(\mathbf{k})$ (blue line), and $V_{+-}(\mathbf{k})$
(green line)
  along the Fermi pocket around $Y$ ($\Gamma$)
 in the $s_{\pm}$-state [panel
(a/c)] and the d-wave  state [panel (b/d)].  (e) and (f): the
  quasiparticle gap on the electron Fermi pocket (red line) and hole Fermi pocket
 (blue line) for $s_{\pm}$- and d-wave states.
}
\end{figure}

Let us examine the pairing strength at the FS around the Fermi pockets
$Y$ and $\Gamma$. For a Fermi wavevector
$\mathbf{k_F}$ on the Fermi pocket centered at $C=(k^c_x, k^c_y)$,
we define an angle $\theta = \arctan (k^F_y - k^c_y) / (k^F_x -
k^c_x)$. The $\theta$-dependences of
 $V(\mathbf{k})$ are plotted in Fig.
~\ref{fig:gap}.  For the  $s_{\pm}$ state, $|V_{++}|>>
|V_{+-}|$, $|V_{--}|$ on $Y$-pocket. This suggests
that the SC pairing is mainly due to the electron pairing of the same orbital.
At the pocket centered at $\Gamma$, $V_{+-}$ is negligibly small,
 so that the SC pairing is mainly due to the hole pairings.
We emphasize that although there are nodal lines,
$V_{++}$ on pocket $Y$ and $V_{--}$ on pocket $\Gamma$ are always finite.
The quasi-particle energy on the Fermi pockets are given by $E_{-}(\mathbf{k})$, which are shown
in Fig. \ref{fig:gap}(e). There is a full gap on both Fermi pockets around $Y$  and $\Gamma$,
consistent with recent ARPES and Andreev reflection spectroscope results.
Because of the above analyses, we have
$E_{-}(\mathbf{k})\approx V_{++}(\mathbf{k})$ around $Y$ and
$E_{-}(\mathbf{k})\approx V_{--}(\mathbf{k})$ around $\Gamma$.
The results for the d-wave state are also shown in Fig.~\ref{fig:gap}.
The nodal line of $V_{--}(\mathbf{k})$ crosses the hole Fermi pocket and leads
to a d-wave like quasiparticle spectrum . The quasiparticle
energy at the nodal point is given by
$E_{\mathbf{k}}= V^2_{+-}(\mathbf{k})/E_{+}(\mathbf{k})$.
Since $V_{+-}(\mathbf{k}) \neq 0$, but small, $E_{\mathbf{k}}$ is non-zero but very tiny
[not distinguishable from 0 in Fig. \ref{fig:gap}(f)].

To better understand the SC pairing and its symmetry found above,
we examine the pair scatterings in the orbital representation 
(intra and inter-orbitals) near the Fermi pockets. 
The spin singlet pairing interaction in $H_2$ can be written as
\begin{eqnarray}
H_{2}^{orb}=-\frac{1}{N}\sum_{\mathbf{k}\mathbf{k'}, nm m'n'
}A_{nm}^{m'n'}(\mathbf{q})\hat{b}^{\dag}_{n
m}(\mathbf{k})\hat{b}^{n' m'}(\mathbf{k'}) \label{EQ:Hsinglet},
\end{eqnarray}
with $\mathbf{q}=\mathbf{k}-\mathbf{k'}$, $ \hat{b}^{n
m}(\mathbf{k})=\frac{1}{\sqrt{2}}\langle \hat{c}_{\mathbf{k}n
\uparrow}\hat{c}_{-\mathbf{k} m\downarrow}-\hat{c}_{\mathbf{k}
n \downarrow}\hat{c}_{-\mathbf{k}m \uparrow} \rangle $. 
$A_{nm}^{m'n'}(\mathbf{q})$ 
is the Fourier transform of
$A_{nm}^{m'n'}(ij)$. $H_{2}^{orb}$ describes the pair scattering processes
between two pairs of electrons with momentum $(\mathbf{k},-\mathbf{k})$
and $(\mathbf{k}',-\mathbf{k'})$. Much of physics may be gained
by examining
the orbital diagonal term
$\hat{b}^{nn}(\mathbf{k})$. 
Denote $\widetilde{A}_{nn'}(\mathbf{q})= A_{nn}^{n'n'}(\mathbf{q})$,
with $\mathbf{q}=(q_x,q_y)$, we find
\begin{eqnarray}
  \label{eq:1}
\widetilde{A}_{11}(\mathbf{q})&=&\frac{4U(t_1^2c_x +t_2^2c_y)}{U^2-J^2}+ (\frac{1}{U+J} +\frac{2}{U-J})4t_3^2c_xc_y\notag \\
\widetilde{A}_{22}(\mathbf{q}) &=& A_{11}(q_y,q_x) \notag \\
\widetilde{A}_{12}(\mathbf{q})&=&\frac{4}{U+J}t_3^2c_xc_y -\frac{4J}{U^2-J^2}t_1t_2(c_x+c_y)
\end{eqnarray}
where $c_x=\cos{q_{x}}$, $c_y=\cos{q_{y}}$, and we have set $t_{4}
=t_{3}$ for simplicity. Since we have small Fermi pockets, the pair scattering wave vectors are
 $\mathbf{q} \approx (0,0)$ within the same pocket , and $\mathbf{q} \approx (\pi,0)$ or $(0,\pi)$ 
between the pockets $\Gamma$ and  $X$ or $Y$, and $\mathbf{q} \approx (\pi,\pi)$
between the pockets $X$ and $Y$, as illustrated in Fig.\ref{fig:fs}.  
From Eq. (8), we find that the intra-pocket pair scatterings are always attractive ($\tilde A(0,0)>0$),
and strongest between the same orbital, and the pair scatterings between hole and 
electron pockets are always repulsive ($\tilde A(0,\pi) <0$). The pair scattering between the two electron
pockets at $X$ and $Y$ points is mainly between two different orbitals, and 
 $\tilde A_{12}(\pi,\pi)$ is attractive at small $J/U$, and repulsive at large $J/U$. 
This qualitatively explains the relative signs in the order parameters
among the different Fermi pockets in both $s_{\pm}$ and d-wave states as shown in
Fig.\ref{fig:pairing}.  The scattering amplitudes in the case $J=0$ are shown in Fig. 1,
which is of $s_{\pm}$-symmetry. 

We have used Eq. \eqref{EQ:Hsinglet} and \eqref{eq:1} to examine the effect to the superconductivity due to the 
simplification of the 2-orbital model, which results in the shift of a hole Fermi pocket
from the $\Gamma$- to $M$- point.  We have found that the qualitative
physics obtained from our study of the 2-orbital model remains the same except the parameter space
for the extended s-wave state is enlarged when more accurate band structure is considered.
To further ensure the qualitative conclusions of our theory, 
we have examined a 3-orbital model as in Ref. \cite{PALee}, in which there are two hole pockets 
around $\Gamma$ in the unfolded BZ, which is better in agreement with the LDA calculations.  
We have extended our analyses of Eq. (7)
to that model and the pairing symmetries are found essentially the same as from the 2-orbital model.

In summary we have examined superconductivity in iron pnictides 
using a 2-orbital Hubbard model at the large U limit. 
An extended s-wave pairing is found most stable in
a large parameter space, consistent with early theories starting with weak coupling (small U) and with
ARPES \cite{Ding} and tunneling experiments \cite{Chien}. Contrary to some of weak coupling theories, we
find that the pairing is mainly from the pair scattering within the 
same Fermi pocket.  Our analyses suggest some
similarities between the superconductivity in iron pnictides and
in the cuprates.  We wish to acknowledge the partial support
from RGC grant of HKSAR and from Swiss National Foundation through
the MANEP network.

\end{document}